\documentclass[showpacs,amsmath,amssymb,twocolumn]{revtex4}
\usepackage{epsfig,dcolumn,bm}

\begin{document}

\title{Dynamical coupled-channel study of $K^*\bar K^*$ and $\omega\phi$ states in a chiral quark model}

\author{W.L. Wang$^{1,2}$}
\author{Z.Y. Zhang$^{1,2}$}
\affiliation
{\small $^1$Institute of High Energy Physics, P.O. Box 918-4, Beijing 100049, China\\
$^2$Theoretical Physics Center for Science Facilities (TPCSF), CAS,
Beijing 100049, China}

\begin{abstract}

A dynamical coupled-channel study of $K^* \bar K^*$ state with isospin 0 and $\omega\phi$ state is performed within both the chiral SU(3) quark model and the extended chiral SU(3) quark model by solving a resonating group method (RGM) equation. The model parameters are taken from our previous work, which gave a satisfactory description of the energies of the octet and decuplet baryon ground states, the binding energy of the deuteron, the nucleon-nucleon ($NN$) scattering phase shifts, and the hyperon-nucleon ($YN$) cross sections. The results show that the interactions of $K^* \bar K^*$ states are attractive, which consequently result in $K^*\bar K^*$ bound states with the binding energies of about $10-70$ MeV, and contrarily, no $\omega\phi$ bound state is obtained. The channel coupling effect of $K^*\bar K^*$ and $\omega\phi$ is found to be considerably large, which makes the binding of $K^*\bar K^*$ $5-45$ MeV deeper. The plausible interpretation of $f_0(1710)$ and $X(1812)$ being $K^*\bar K^*$ dominated states is briefly discussed.

\end{abstract}


\maketitle

\section{Introduction}

Since its discovery in 1982 \cite{edwards}, the $f_0(1710)$ has attracted a lot of discussions and debates on its structures \cite{dooley91,sexton95,wang00,chen01,gengls10f0}. In literature, it has ever been explained as glue-ball \cite{sexton95}, $K \bar K$ excited state \cite{wang00,chen01}, a linear composition of $K^*\bar K^*$ and $\omega\phi$ \cite{dooley91}, {\it et al.}. In Ref.~\cite{gengls10f0}, the authors performed a systematic study of vector meson-vector meson interaction using a chiral unitary approach, and the $f_0(1710)$ is found to be a dynamically generated resonance dominated by $K^*\bar K^*$ state.

An $\omega\phi$ near-threshold enhancement has been reported by the BES-II Collaboration in 2006 in $J/\psi \to \gamma X \to \gamma \omega \phi$ reaction \cite{ablikim}. The partial wave analysis shows that this state, called $X(1812)$, favors $J^P=0^+$, and its mass and width are $M=1812^{+19}_{-26}\pm18$ MeV and $\Gamma=105 \pm 20 \pm28$ MeV. There are many explanations for this state in literature \cite{li06,buccella07,chao06,bicudo07,bugg06,he06,rosner06,zhang07,chen07,zhu08},
such as tetraquark state \cite{li06,buccella07}, hybrid \cite{chao06}, glue-ball \cite{bicudo07}, {\it et al.} (See Ref.~\cite{zhu08} for a review). Recently the Belle Collaboration has also performed a search for $X(1812)$ in the decay $B^\pm \to K^\pm \omega\phi$, but no evidence for the existence of $X(1812)$ state has been reported \cite{liu09}.

In Ref.~\cite{zhao06}, the authors has examined the intermediate meson re-scattering contributions to the $J/\psi \to \gamma X \to \gamma \omega \phi$ process by assuming that $X=f_0(1710)$ with a mass at $1.74-1.81$ GeV. It is found that the contributions from $K^*\bar K^*$ re-scattering can produce some enhancement near the $\omega\phi$ threshold, while the other intermediate meson re-scatterings, such as $K\bar K$ and $\kappa \bar \kappa$ re-scatterings, are found to be small.

Since the structures of $f_0(1710)$ and $X(1812)$ are still unclear, and moreover, from the points of view of Refs.~\cite{gengls10f0,zhao06}, the structures of both $f_0(1710)$ and $X(1812)$ are related to $K^* \bar K^*$ and $\omega\phi$ states, it would be interesting and helpful to perform a dynamical coupled-channel study of the $K^* \bar K^*$ and $\omega\phi$ interactions within an approach other than the chiral unitary approach as used in Ref.~\cite{gengls10f0}.

The chiral SU(3) quark model and the extended chiral SU(3) quark model have been widely used in the past few years and considerable achievements have been made in studying the hadron-hadron interactions \cite{zyzhang97,fhuang04kn,fhuang04nkdk,lrdai03,fhuang05kne,fhuang05lksk,
fhuang05dklksk,fhuang07kbn,fhuang08kbn,fhuang06nphi,wlwang08xikb,wlwang07omepi}. In the chiral SU(3) quark model, the quark-quark interaction contains confinement, one-gluon exchange (OGE) and boson exchanges stemming from scalar and pseudoscalar nonets, and the short range quark-quark interaction is found to be provided by OGE and quark exchange effects. In the extended chiral SU(3) quark model, the coupling of the quark and vector meson fields is included, and the OGE that plays an important role in the short-range quark-quark interaction in the original chiral SU(3) quark model is now nearly replaced by the vector meson exchanges (VMEs). In other words, the short range interaction mechanisms are quite different in these two models. During the past few years, the chiral SU(3) quark model and the extended chiral SU(3) quark model have been quite successful in reproducing the energies of the baryon ground states, the binding energy of deuteron, the $NN$ and $KN$ scattering phase shifts, and the $YN$ cross sections \cite{zyzhang97,fhuang04kn,fhuang04nkdk}. When applied to the systems of $NN$, $\Delta K$, $\Lambda K$ and $\Sigma K$ \cite{lrdai03,fhuang05kne,fhuang05dklksk}, those two models are found to give similar results and thus OGE or VMEs for short range interaction mechanism is indistinguishable. Recently in a preliminary combined study of $KN$ and $\bar{K}N$ interactions it is found that those two models might give different contributions \cite{fhuang07kbn,fhuang08kbn}. Based on these achievements we have obtained, it is interesting to study more hadron-hadron systems within the chiral SU(3) quark model and the extended chiral SU(3) quark model in order to get more information about the shout range quark-quark interaction mechanisms and to see how far we can go with these two models.

In this work, we perform a dynamical coupled-channel study of the $K^*\bar K^*$ and $\omega\phi$ states with isospin $I=0$ and spin-parity $J^P=0^+$, $1^+$ and $2^+$ within both the chiral SU(3) quark model and the extended chiral SU(3) quark model by solving a RGM equation. All the model parameters are taken from our previous studies of $\Delta K$, $\Lambda K$, $\Sigma K$, $N\phi$, $\Xi \bar K$, and $\Omega\pi$ systems \cite{fhuang05dklksk,fhuang06nphi,wlwang08xikb,wlwang07omepi}, and we don't have any free adjustable parameters here. Our results show that the interactions of $K^* \bar K^*$ are attractive, and in spin $S=0$ channel a bound state is obtained in the extended chiral SU(3) quark model with the energy of about $1720-1727$ MeV, which is consistent with the explanations that $f_0(1710)$ is a dominated $K^*\bar K^*$ state, similar to the results from Ref.~\cite{gengls10f0}, and the $X(1812)$ observed in $J/\psi \to \gamma X \to \gamma \omega \phi$ might be mainly from the effects of the tail of $f_0(1710)$ through $K^*\bar K^*$ re-scattering process, as pointed out in Ref.~\cite{zhao06}. The results for $K^*\bar K^*$ states with spin $S=1$ and $S=2$ are also shown and discussed.

The paper is organized as follows. In the next section, the framework of the chiral SU(3) quark model and the extended chiral SU(3) quark model are briefly introduced. The results for the $K^*
\bar K^*$ and $\omega\phi$ states are shown and discussed in Sec. III. Finally, the summary is given in Sec. IV.

\section{Formulation}

Both the chiral SU(3) quark model and the extended chiral SU(3)
quark model have been widely described in the literatures
\cite{fhuang04kn,fhuang05kne,fhuang05dklksk}, and we refer the
reader to those works for details. Here we just show the salient
features.

In both models, the total Hamiltonian of meson-meson systems can be
written as
\begin{equation}
H=\sum_{i}T_{i}-T_{G}+V_{13}+V_{\bar 2\bar 4}+\sum_{i,j}V_{i\bar j},
\end{equation}
where $T_G$ is the kinetic energy operator for the center-of-mass
motion, and $V_{13}$, $V_{\bar 2\bar 4}$ and $V_{i\bar j}$ represent
the quark-quark, antiquark-antiquark and quark-antiquark
interactions, respectively. $V_{13}$ is expressed as
\begin{equation}
V_{13}= V^{\rm OGE}_{13} + V^{\rm conf}_{13} + V^{\rm ch}_{13},
\end{equation}
where $V_{13}^{\rm OGE}$ is the OGE interaction and $V_{13}^{\rm
conf}$ the confinement potential. $V^{\rm ch}_{13}$ represents the
effective quark-quark potential induced by one-boson exchanges. In
our original chiral quark model, $V^{\rm ch}_{13}$ includes the
scalar boson exchanges and the pseudoscalar boson exchanges,
\begin{eqnarray}
V^{\rm ch}_{13} = \sum_{a=0}^8 V_{\sigma_a}({\bm
r}_{ij})+\sum_{a=0}^8 V_{\pi_a}({\bm r}_{ij}),
\end{eqnarray}
and when the model is extended to include the vector boson
exchanges, $V^{\rm ch}_{13}$ can be written as
\begin{eqnarray}
V^{\rm ch}_{13} = \sum_{a=0}^8 V_{\sigma_a}({\bm
r}_{ij})+\sum_{a=0}^8 V_{\pi_a}({\bm r}_{ij})+\sum_{a=0}^8
V_{\rho_a}({\bm r}_{ij}).
\end{eqnarray}
Here $\sigma_{0},...,\sigma_{8}$ are the scalar nonet fields,
$\pi_{0},..,\pi_{8}$ the pseudoscalar nonet fields, and
$\rho_{0},..,\rho_{8}$ the vector nonet fields. The expressions of
these potentials can be found in the literatures
\cite{fhuang04kn,fhuang05kne,fhuang05dklksk}.

$V_{\bar 2\bar 4}$ in Eq.~(1) represents the antiquark-antiquark
interaction,
\begin{equation}
V_{\bar 2\bar 4}= V^{\rm OGE}_{\bar 2\bar 4} + V^{\rm conf}_{\bar
2\bar 4} + V^{\rm ch}_{\bar 2\bar 4},
\end{equation}
where $V^{\rm OGE}_{\bar 2\bar 4}$ and $V^{\rm conf}_{\bar 2\bar 4}$
are obtained by replacing ${\bm\lambda}^c_1\cdot{\bm\lambda}^c_3$ in
Eqs.~(3) and (4) with ${\bm\lambda}^{c*}_{\bar
2}\cdot{\bm\lambda}^{c*}_{\bar 4}$, and $V^{\rm ch}_{\bar 2\bar 4}$
is in the same form as $V^{\rm ch}_{12}$.

$V_{i\bar j}$ in Eq.~(1) represents the quark-antiquark interaction,
\begin{equation}
V_{i\bar j}= V^{\rm OGE}_{i\bar j} + V^{\rm conf}_{i\bar j} + V^{\rm
ch}_{i\bar j},
\end{equation}
where $V^{\rm OGE}_{i\bar j}$ and $V^{\rm conf}_{i\bar j}$ are
obtained by replacing the ${\bm\lambda}^c_1\cdot{\bm\lambda}^c_3$ in
Eqs.~(3) and (4) with
$-{\bm\lambda}^{c}_{i}\cdot{\bm\lambda}^{c*}_{\bar j}$, and
$V_{i\bar{j}}^{\rm ch}$ can be obtained from the G parity
transformation:
\begin{equation}
V_{i\bar{j}}^{\rm ch}=\sum_{k}(-1)^{G_k}V_{ij}^{{\rm ch},k},
\end{equation}
with $(-1)^{G_k}$ being the G parity of the $k$th meson.

All the model parameters are taken from our previous works
\cite{fhuang05dklksk,fhuang05lksk,fhuang06nphi,wlwang08xikb,wlwang07omepi} and
their values are listed in Table~I, where the first set is for the
original chiral SU(3) quark model, the second and third sets are for
the extended chiral SU(3) quark model by taking $f_{\rm chv}/g_{\rm
chv}$ as $0$ and $2/3$, respectively. Here $g_{\rm chv}$ and $f_{\rm
chv}$ are the coupling constants for vector coupling and tensor
coupling of the vector meson fields, respectively. $b_u$ is the
harmonic-oscillator width parameter, and $m_{u(d)}$ and $m_s$ the
$u(d)$ quark and $s$ quark masses. $g_u$ and $g_s$ are the OGE
coupling constants and $g_{\rm ch}$ the coupling constant for scalar
and pseudo-scalar chiral field coupling. $m_\sigma$ is the mass for
$\sigma$ meson and $a^c$ represents the strength of the confinement
potential. All these three sets of parameters can give a
satisfactory description of the masses of the baryon ground states,
the binding energy of deuteron, and the $NN$ scattering phase
shifts.

{\small
\begin{table}[tb]
\caption{\label{para} Model parameters. The meson masses and the
cutoff masses: $m_{\sigma'}=980$ MeV, $m_{\kappa}=980$ MeV,
$m_{\epsilon}=980$ MeV, $m_{\pi}=138$ MeV, $m_K=495$ MeV,
$m_{\eta}=549$ MeV, $m_{\eta'}=957$ MeV, $m_{\rho}=770$ MeV,
$m_{K^*}=892$ MeV, $m_{\omega}=782$ MeV, $m_{\phi}=1020$ MeV, and
$\Lambda=1100$ MeV.}
\begin{tabular}{cccc}
\hline\hline
  & $\chi$-SU(3) QM & \multicolumn{2}{c}{Ex. $\chi$-SU(3) QM}  \\
  &   I   &    II    &    III \\  \cline{3-4}
  &  & $f_{chv}/g_{chv}=0$ & $f_{chv}/g_{chv}=2/3$ \\
\hline
 $b_u$ (fm)  & 0.5 & 0.45 & 0.45 \\
 $m_u$ (MeV) & 313 & 313 & 313 \\
 $m_s$ (MeV) & 470 & 470 & 470 \\
 $g_u^2$     & 0.766 & 0.056 & 0.132 \\
 $g_s^2$     & 0.846 & 0.203 & 0.250 \\
 $g_{ch}$    & 2.621 & 2.621 & 2.621  \\
 $g_{chv}$   &       & 2.351 & 1.973  \\
 $m_\sigma$ (MeV) & 595 & 535 & 547 \\
 $a^c_{uu}$ (MeV/fm$^2$) & 46.6 & 44.5 & 39.1 \\
 $a^c_{us}$ (MeV/fm$^2$) & 58.7 & 79.6 & 69.2 \\
 $a^c_{ss}$ (MeV/fm$^2$) & 99.2 & 163.7 & 142.5 \\
 $a^{c0}_{uu}$ (MeV)  & $-$42.4 & $-$72.3 & $-$62.9 \\
 $a^{c0}_{us}$ (MeV)  & $-$36.2 & $-$87.6 & $-$74.6 \\
 $a^{c0}_{ss}$ (MeV)  & $-$33.8 & $-$108.0 & $-$91.0 \\
\hline\hline
\end{tabular}
\end{table}}

With all parameters determined, the $K^* \bar K^*$ state can be
dynamically studied in the framework of the RGM. The wave function
of the $K^* \bar K^*$ system is of the form
\begin{eqnarray}
\Psi={\cal A}[{\hat \psi}_{K^*}(\bm \xi_1) {\hat \psi}_{\bar
K^*}(\bm \xi_2) \chi({\bm R}_{K^* \bar K^*})],
\end{eqnarray}
where ${\bm \xi}_1$ is the internal coordinates for the cluster
$K^*$, and ${\bm \xi}_2$ the internal coordinate for the cluster
$\bar K^*$. ${\bm R}_{K^* \bar K^*}\equiv {\bm R}_{K^*}-{\bm
R}_{\bar K^*}$ is the relative coordinate between the two clusters,
$K^*$ and $\bar K^*$. The ${\hat \psi}_{K^*}$ is the internal wave
function of $K^*$, as well as ${\hat \psi}_{\bar K^*}$ is the
internal wave function of $\bar K^*$, and $\chi({\bm R}_{K^* \bar
K^*})$ is the relative wave function of the two clusters. The symbol
$\cal A$ is the anti-symmetrizing operator defined as
\begin{equation}
{\cal A}\equiv(1-P_{13})(1+P_{K^* \bar K^*}).
\end{equation}
Expanding unknown $\chi({\bm R}_{K^* \bar K^*})$ by employing
well-defined basis wave functions, such as Gaussian functions, one
can solve the RGM equation for a bound-state problem or a scattering
one to obtain the binding energy or scattering phase shifts for the
two-cluster systems. The details of solving the RGM equation can be
found in Refs.~\cite{wildermuth77,kamimura77,oka81}.

\section{Results and discussions}

\begin{figure}[tb]
\epsfig{file=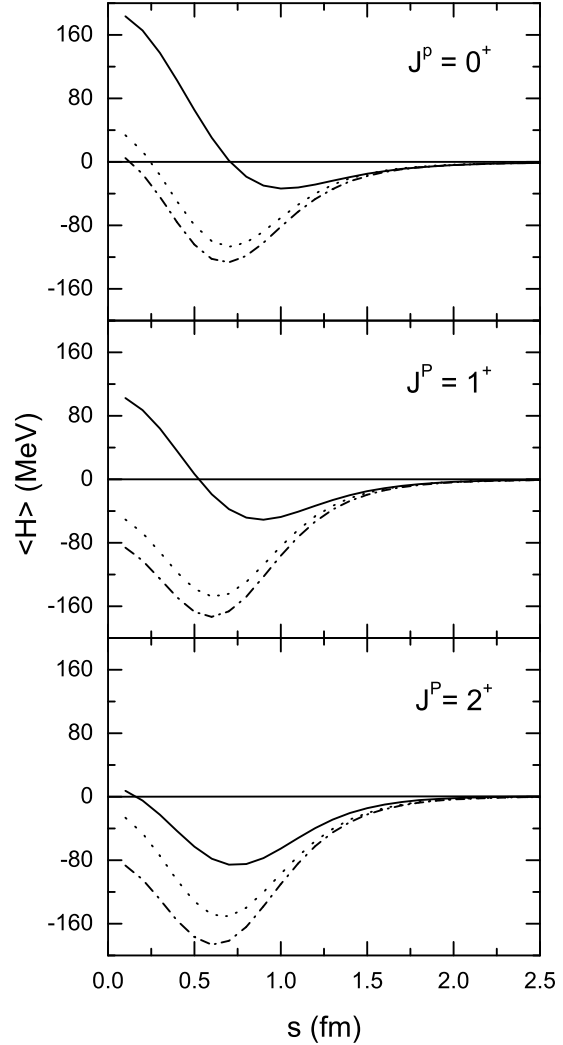,width=0.45\textwidth} \caption{\small The GCM
matrix elements of the Hamiltonian for $K^*\bar K^*$. The solid,
dash-dotted and dotted lines represent the results obtained in models
I, II and III, respectively.}
\end{figure}

As discussed in the Introduction, the structures of $f_0(1710)$ and $X(1812)$ are still unclear, but from the points of view of Refs.~\cite{gengls10f0,zhao06}, they both are related to $K^* \bar K^*$ and $\omega\phi$ states. It is straightforward that a dynamical coupled-channel study of the $K^* \bar K^*$ and $\omega\phi$ states within an approach other than the chiral unitary approach as used in Ref.~\cite{gengls10f0} would be interesting and helpful with the understanding of the structures of $f_0(1710)$ and $X(1812)$.

The chiral SU(3) quark model and the extended chiral SU(3) quark model have been quite successful in describing the data for $NN$, $YN$, $KN$, and $\bar KN$ scattering processes in the past few years \cite{zyzhang97,lrdai03,fhuang04kn,fhuang04nkdk,fhuang05kne,fhuang07kbn,fhuang08kbn}. In the present work, we perform a RGM dynamical study of the $S$-wave $K^* \bar K^*$ systems with isospin $I=0$ and spin-parity $J^P=0^+$, $1^+$ and $2^+$ within our chiral quark models. All the model parameters are taken from our previous work \cite{fhuang05dklksk,fhuang05lksk,fhuang06nphi,wlwang08xikb,wlwang07omepi}, and with which a good description of the energies of octet and decuplet baryon ground states, the binding energy of deuteron, and the $NN$ scattering phase shifts has been achieved \cite{zyzhang97,lrdai03}.

First we would like to show and discuss the results from a single-channel calculation. Figure~1 shows the results of diagonal matrix elements of the Hamiltonian for $K^* \bar K^*$ systems with spin-parity $J^P=0^+$, $1^+$ and $2^+$ in the generator coordinate method (GCM) \cite{wildermuth77} calculation, which can be regarded as the effective Hamiltonian of two clusters $K^*$ and $\bar K^*$ qualitatively. In Fig.~1, $H$ includes the kinetic energy of $K^*\bar K^*$ relative motion and the effective potential between $K^*$ and $\bar K^*$, and $s$ denotes the generator coordinate which can qualitatively describe the distance between the two clusters. The solid, dash-dotted and dotted lines represent results calculated by using those three sets of parameters as depicted in Table~1. From Fig.~1, one sees that the $K^* \bar K^*$ interaction is always attractive in the medium and long range for all spin cases and for all those three sets of parameters. One also sees that the attraction is weakest in Model I, i.e. the chiral SU(3) quark model, while the attractions are much stronger in Models II and III, i.e. the extended chiral SU(3) quark model with the ratio of tensor coupling and vector coupling for vector meson fields $f_{\rm chv}/g_{\rm chv}=0$ and $2/3$ respectively. This is simply because in the extended chiral SU(3) quark model, the VME has been included, which provides additional attraction from $\rho$-exchange other than the attraction from $\sigma$ exchange in the original chiral SU(3) quark model.

{\small
\begin{table}[tb]
\caption{The binding energy of $K^*\bar K^*$ with isospin $I=0$ in one-channel calculation (MeV).}
\begin{tabular*}{85mm}{@{\extracolsep\fill}lccc}
\hline\hline
 Model & $S=0$ &  $S=1$  &  $S=2$ \\
\hline
I  & $-$ & $-$  & $11$ \\
II  & $28$ & $61$  & $71$ \\
III  & $16$ & $42$  & $44$ \\
\hline\hline
\end{tabular*}
\end{table}}

In order to see whether the $K^*\bar K^*$ attraction can result in a bound state or not, we have
solved the RGM equation for a bound state problem. The binding energies of $K^*\bar K^*$ for all spin cases in all the models are shown in Table~II. One sees that in model I, i.e. the original chiral SU(3) quark model, the $K^*\bar K^*$ states are unbound for both spin $S=0$ and $S=1$ channels due to the insufficient attractive interaction, while in models II and III, i.e. the extended chiral SU(3) quark model with $f_{\rm chv}/g_{\rm chv}=0$ and $f_{\rm chv}/g_{\rm
chv}=2/3$, the $K^*\bar K^*$ states are bound with the binding energies of about $28$ and $16$ MeV for $S=0$ and $61$ and $42$ MeV for $S=1$, respectively.  The $K^* \bar K^*$ states with spin $S=2$ are always bound in all those three models, with the binding energies of about 11$-$71 MeV. The above binding information can be qualitatively understood from Fig.~1. There, one sees: (1) the $K^*\bar K^*$ attractions in models II and III are much stronger that those in model I, and (2) the attractions of $K^*\bar K^*$ with spin $S=2$ are much stronger that those with spin $S=0$ and $S=1$.

Further analysis shows that the $K^*\bar K^*$ interaction is dominated by $\sigma$, $\pi$ and $\rho$ exchanges. For spin $S=0$ and $S=1$ cases, in the chiral SU(3) quark model, $\sigma$ exchange provides strong attraction and $\pi$ exchange provides strong repulsion, and the sum of these two is weakly attractive. Thus, no $K^*\bar K^*$ bound state can be obtained due to the insufficiency of the attraction. In the extended chiral SU(3) quark model, the coupling of quark and vector meson fields is included, which provides additional strong attraction from $\rho$ exchange other than the contributions from $\sigma$ and $\pi$ exchanges. As a consequence, spin $S=0$ and $S=1$ $K^*\bar K^*$ bound states are obtained in this model since now the strengths of the attractions are sufficiently strong. For spin $S=2$ case, the $\pi$ exchange provides attraction for $K^*\bar K^*$ interactions, instead of repulsion as in the spin $S=0$ and $S=1$ cases. As a result, $K^*\bar K^*$ bound states are obtained in all the models and the corresponding binding energies are considerably larger than those for $K^*\bar K^*$ states with spin $S=0$ and $S=1$.

\begin{figure}[htb]
\epsfig{file=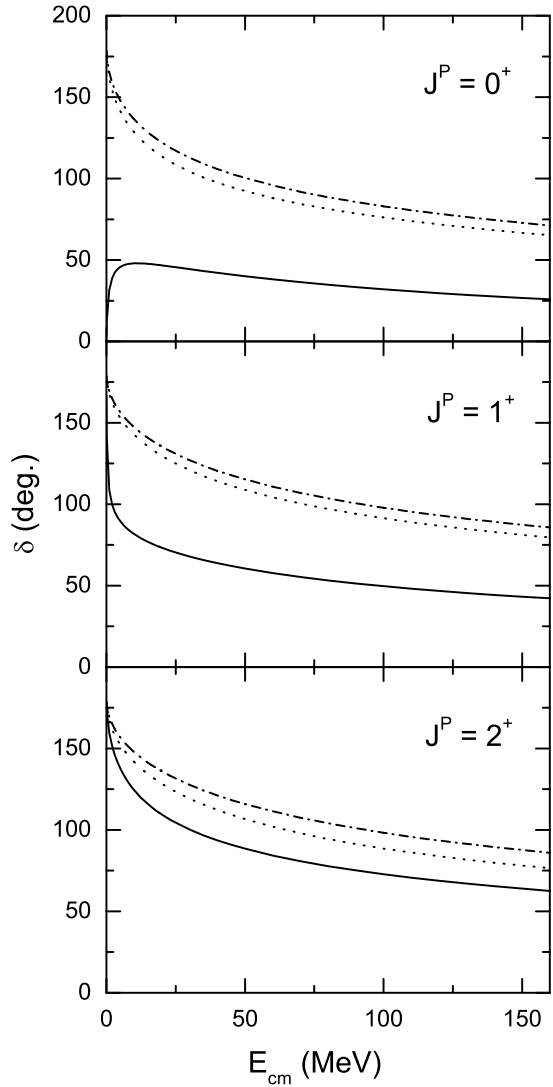,width=0.45\textwidth} \caption{\small
$K^* \bar K^*$ $S$-wave phase shifts as a function of the energy of
the center of mass motion in the one-channel calculation. Same
notation as in Fig.~1.}
\end{figure}

The study of $K^* \bar K^*$ elastic scattering processes has also been performed by solving the RGM equation. The calculated $S$-wave $K^* \bar K^*$ phase shifts are shown in Fig.~2 as a function of $K^* \bar K^*$ center of mass energy subtracted by the $K^*\bar K^*$ threshold energy. One sees that the phase shifts are always positive, denoting attractive $K^* \bar K^*$ interactions, in different models and spin channels. One also sees that the magnitude of the phase shifts are higher for all spin channels in the extended chiral SU(3) quark model than those in the chiral SU(3) quark model, which indicates the more attractive interaction in the extended chiral SU(3) quark model. Note that in the chiral SU(3) quark model the magnitude of the phase shifts for spin $S=2$ channel is much higher than those for spin $S=0$ and $S=1$ channels, which indicates that the interaction of $K^* \bar K^*$ with spin $S=2$ is much more attractive than those with $S=0$ and $S=1$. We mention that all these information from Fig.~1, Table~II and Fig.~2 about the $K^*\bar K^*$ interaction properties are consistent with each other.

\begin{figure}[tb]
\epsfig{file=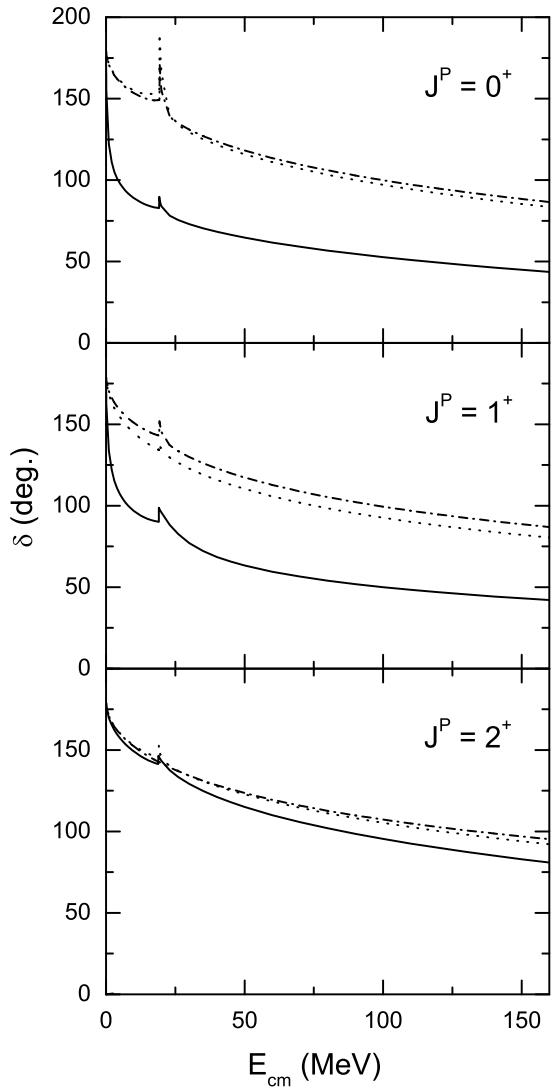,width=0.45\textwidth} \caption{\small
$K^* \bar K^*$ $S$-wave phase shifts as a function of the energy of
the center of mass motion in the coupled-channel calculation. Same
notation as in Fig.~1.}
\end{figure}

{\small
\begin{table}[tb]
\caption{The binding energy of $K^*\bar K^*$ with isospin $I=0$ in coupled-channel calculation (MeV).}
\begin{tabular*}{85mm}{@{\extracolsep\fill}lccc}
\hline\hline
 Model & $S=0$ &  $S=1$  &  $S=2$ \\
\hline
I  & $-$ & $-$  & $55$ \\
II  & $64$ & $69$  & $100$ \\
III  & $57$ & $46$  & $89$ \\
\hline\hline
\end{tabular*}
\end{table}}

In Ref.~\cite{wlwang10omefi} we have studied the $\omega\phi$ system and the results show that although the $\omega\phi$ interaction is attractive in intermediate and long ranges, no $\omega\phi$ bound state or resonance state can be obtained within our chiral quark models.

Since the threshold of $\omega\phi$ is only $18$ MeV higher than that of $K^*\bar K^*$, the channel-coupling effect of these two channels is expected to be large and unnegligible. This effect is investigated here by dynamically solving coupled-channel RGM equations for the scattering problem and for the bound state problem. The calculated $K^*\bar K^*$ scattering phase shifts in each channel and each model are shown in Fig.~3, and the corresponding $K^*\bar K^*$ binding energies are shown in Table III. Comparing Fig.~3 with Fig.~2, we see that for spin $S=0$ and $S=2$ cases the channel coupling of $K^* \bar K^*$ and $\omega\phi$ makes the $K^*\bar K^*$ phase shifts get much higher in the magnitude, while for spin $S=1$ case, the $K^*\bar K^*$ and $\omega\phi$ channel coupling does not cause visible effects on the $K^*\bar K^*$ phase shifts. Based on this one may expect that when the channel coupling of $K^*\bar K^*$ and $\omega\phi$ is considered, for spin $S=0$ and $S=2$ cases the $K^*\bar K^*$ binding energies will get much bigger than those from a single-channel calculation, while for spin $S=1$ case, the $K^*\bar K^*$ binding energy will not change too much. These features are clearly manifested by the numerical numbers of the $K^*\bar K^*$  binding energies from a coupled-channel calculation as shown in Table III, where a $5-45$ MeV deeper binding energy is shown as coming from the $K^*\bar K^*$ and $\omega\phi$ channel coupling effects.

In Ref.~\cite{gengls10f0}, the vector meson-vector meson interaction has been studied on the hadron level within the chiral unitary approach. In isospin $I=0$ and spin $S=0$ channel, a dynamically generated resonance with the mass around $1726$ MeV is reported. This state is dominated by $K^*\bar K^*$ state. In our calculation, we see from Table III that in the chiral SU(3) quark model where the short range quark-quark interaction is dominated by OGE, the $K^*\bar K^*$ state is unbound, while in the extended chiral SU(3) quark model where the short range quark-quark interaction is dominated by VME, the $K^*\bar K^*$ is bound with a corresponding $K^*\bar K^*$ energy of about $1720-1727$ MeV. This means that the attractions stemming from the VMEs play an important role in the forming of the spin $S=0$ $K^*\bar K^*$ bound state, which is similar to the results from Ref.~\cite{gengls10f0}, where the $K^*\bar K^*$ interactions are dominated by VMEs. Of great interest is that we get similar energies for the spin $S=0$ $K^*\bar K^*$ states even using different methods, i.e. constituent quark model and chiral unitary approach.

In Ref.~\cite{gengls10f0}, the isospin $I=0$ and spin $S=0$ $K^*\bar K^*$ state is identified by $f_0(1710)$ since the experimental values of both the mass and the width of $f_0(1710)$ are consistent with their theoretical values. The branching ratios are also discussed there. In our present work, the obtained mass of the isospin $I=0$ and spin $S=0$ $K^*\bar K^*$ state is really consistent with the experimental mass of $f_0(1710)$. We would need to study the decay widths and the branching ratios for this state in future for further identification of this state with $f_0(1710)$.

In Ref.~\cite{zhao06}, in order to understand the $\omega\phi$ near threshold enhancement, i.e. the so-called $X(1812)$, the authors has examined the intermediate meson re-scattering contributions to the $J/\psi \to \gamma X \to \gamma \omega \phi$ process by assuming that $X=f_0(1710)$ with a mass at $1.74-1.81$ GeV. It is found that the contributions from $K^*\bar K^*$ re-scattering do produce some enhancement near the $\omega\phi$ threshold. Our previous work \cite{wlwang10omefi} and the present work show that there is no $\omega\phi$ bound state or resonance state, but there is an isospin $I=0$ and spin $S=0$ $K^*\bar K^*$ bound state with the energy of about $1720-1727$ MeV, which couples relatively strong to $\omega\phi$. This provides some sort of support for the $X(1812)$ observed in $\omega\phi$ channel being from the effects of the tail of $f_0(1710)$ through $K^*\bar K^*$ re-scattering process, as pointed out in Ref.~\cite{zhao06}. Future work about the decay processes needs to be done for further clarification.

For isospin $I=0$ and spin $S=2$ channel, in Ref.~\cite{gengls10f0}, the dynamically generated resonance with the mass around $1525$ MeV is reported. This state couples to $K^*\bar K^*$, $\omega\phi$, $\phi\phi$, $\rho\rho$ and $\omega\omega$ channels, and is identified with the $f_2'(1525)$ resonance. In the present work, our results listed in Table III show that in this channel the $K^*\bar K^*$ states are bound with the energies of about $1684-1729$ MeV. The $\phi\phi$, $\rho\rho$ and $\omega\omega$ channels need to be included in our next-step work for a further comparison with the results from Ref.~\cite{gengls10f0}.

For isospin $I=0$ and spin $S=1$ $K^*\bar K^*$ system, our results listed in Table III show that there is no bound state in chiral SU(3) quark model, but there are bound states in the extended chiral SU(3) quark model with the corresponding $K^*\bar K^*$ energies being $1715-1738$ MeV. In Ref.~\cite{gengls10f0}, the pole position in this channel is found to be at $1802-i39$ MeV, which is slightly above the $K^*\bar K^*$ threshold. Investigations from other approaches will be needed to further test the model dependence of the results for this channel.

\begin{figure}[tb]
\epsfig{file=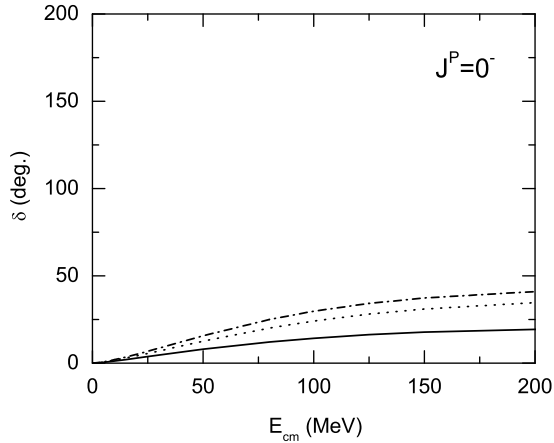,width=0.45\textwidth} \caption{\small $K^* \bar K^*$ $P$-wave phase shifts as a function of the energy of the center of mass motion. Same notation as in Fig.~1.}
\end{figure}

In Ref.~\cite{bes2000}, BES collaboration has reported a broad $0^-$ resonance with mass $M=1800$ MeV and width $\Gamma=500\pm200$ MeV on the decay $J/\psi \rightarrow \gamma K^* \bar K^*$. Here we have performed a study of the $P$-wave $K^* \bar K^*$ states. The calculated phase shifts are shown in Fig.~4. One sees that although positive, the phase shifts are relatively small in the magnitude, which indicates that the $P$-wave $K^*\bar K^*$ interaction is attractive but the strength of the attraction is very week. Further we have solved the RGM equation for bound state problem, and the results show that there are no $K^* \bar K^*$ $P$-wave bound state or resonance state in our model.

\section{Summary}

In this work, we have performed a dynamical coupled-channel study of the $K^* \bar K^*$ and $\omega\phi$ states with isospin $I=0$ and spin $S=0$, $1$, and $2$ in the chiral SU(3) quark model as well as in the extended chiral SU(3) quark model by solving the RGM equation. All the model parameters are taken from our previous work, which can give a satisfactory description of the energies of the baryon ground states, the binding energy of the deuteron, and the $NN$ scattering phase shifts \cite{zyzhang97,lrdai03}. The calculated results show that the interactions of $K^* \bar K^*$ are attractive, and in spin $S=0$ channel a bound state is obtained in the extended chiral SU(3) quark model with the energy of about $1720-1727$ MeV, which is consistent with the explanations that $f_0(1710)$ is a dominated $K^*\bar K^*$ state, similar to the results from Ref.~\cite{gengls10f0}, and the $X(1812)$ observed in $J/\psi \to \gamma X \to \gamma \omega \phi$ might be mainly from the effects of the tail of $f_0(1710)$ through $K^*\bar K^*$ re-scattering process, as pointed out in Ref.~\cite{zhao06}. The $K^*\bar K^*$ bound states with spin $S=1$ and $S=2$ are also obtained. More studies about the decay properties will be performed in our future work for further clarification of these $K^*\bar K^*$ states.

\begin{acknowledgments}
The authors thank Professors B.S. Zou and Q. Zhao for their
helpful discussions. This project was supported by Ministry of
Science and Technology of China (Grant No. 2009CB825200) and China
Postdoctoral Science Foundation (Grant No. 20100480468).
\end{acknowledgments}

\end{document}